# On the possibility of mid-IR supercontinuum generation in As-Se-Te/As-S core/clad fibers with all-fiber femtosecond pump source

E.A. Anashkina[1,*], V.S. Shiryaev[2], G.E. Snopatin[2], S.V. Muraviev[1], and A.V. Kim[1]

[1]Institute of Applied Physics of the Russian Academy of Sciences, Nizhny Novgorod, Russia
[2]G.G. Devyatykh Institute of Chemistry of High-Purity Substances of the Russian Academy of Sciences, Nizhny Novgorod, Russia
*elena.anashkina@gmail.com

**Highlights**:
- Mid-IR SC generation in As-Se-Te/As-S fibers pumped at 2 μm is studied numerically.
- Simulated spectra extending from ~1 μm to more than 8 μm are demonstrated.
- Low-loss As-Se-Te/As-S step-index fibers are manufactured.

**Abstract**
We propose and optimize theoretically a supercontinuum (SC) laser source in the mid-IR based on using As-Se-Te/As-S core/clad step-index fibers and a femtosecond all-fiber laser system at 2 μm. Numerically simulated spectra extending from ~1 μm to more than 8 μm are demonstrated for pump energy of order 100 pJ in a fiber with a core diameter of 2 μm. To the best of our knowledge, the possibility of such long-wavelength spectral conversion of pump pulses at the wavelength of 2 μm in optical fibers is demonstrated for the first time. The theoretical calculations are performed on the base of real low loss step-index As-Se-Te/As-S glass fibers with various core-clad diameter ratios.

**Introduction**
The mid-infrared (IR) ultra-broadband coherent light sources have important applications in remote sensing, biophotonics, homeland security, and so on. The photonic technologies based on mid-IR fibers providing these properties are very promising [1]. Chalcogenide glasses (ChG) have the broadest transmittance windows and the highest third-order nonlinear refractive indices ($n_2$) among all optical glasses, ~200-1000 times higher than that for silica [1, 2]. These characteristics make them ideal candidates for mid-IR nonlinear fiber optics where short sample lengths or ultralow pulse energies are sufficient to elicit nonlinear optical behavior. SC generation in ChG fibers with various compositions, geometrical design, and pumping wavelengths are constantly reported (see [1] and references therein). To date, spectral SCs in the 1.4-13.3 μm and 2-15.1 μm ranges have been demonstrated in As-Se core step-index fibers pumped by optical parametric amplifiers (OPA) at 6.3 μm [3] and at 9.8 μm [4], respectively. An As-Se commercial fiber has allowed generating SC in the 3-8 μm range using an erbium-doped $ZrF_4$-based in-amplifier SC source spanning from 3 to 4.2 μm [5]. SC spectrum spanning 1.5-14 μm has been achieved by pumping a Ge-As-Se-Te fiber at 4.5 μm [6]. An As-S-Se/As-S fiber pumped at 4.8 μm has been used to generate the spectrum in the 1-5 μm range [7]. SCs spanning 1.7-7.5 μm and 0.9-9 μm has been obtained with microstructured As-Se fibers pumped by OPA at 4.4 μm [8] and by SC from ZBLAN fiber in the 0.9-4.1 μm range respectively [9]. SC with red boundary beyond 5 μm has been reported for all-solid microstructured As-Se/As-S fiber pumped at 3.4 μm by OPA [10].

However, despite the substantial efforts which have led to startling results over the past several decades, the full potential of ChG fibers has not been achieved yet [2]. For instance, in the majority of the works devoted to mid-IR SC generation, the pump source initially operates in the mid-IR. But it would be great to start from the standard robust fiber laser systems in the near-IR in order to simplify a scheme and make a photonic device more suitable and convenient for applications. To the best of our knowledge, the reached red boundary of the SC spectra in ChG-based fibers pumped by fiber laser systems in the range of 1.5-2 μm is shorter than 4 μm [11-13]. But our numerical study demonstrates a possibility of SC generation with red boundary beyond 8 μm in As-Se-Te/As-S core/cladding step-index fibers pumped by a femtosecond all-fiber laser system at 2 μm. Such a kind of laser system at 2 μm built on standard telecom components and Er-doped and Tm/Yb co-doped silica fibers has been previously used by us for SC generation in the 2-3 μm range in a germano-silicate fiber [14] and for Raman soliton shifting up to 2.65 μm in tellurite fibers [15]. Here we have synthesized experimentally high-purity $As_{39.4}Se_{55.3}Te_{5.3}$ glass for core and $As_{39.4}S_{60.6}$ glass for cladding and manufactured low-loss fibers with various diameters that can be pumped by this all-fiber system for SC generation. We study numerically nonlinear dynamics of the ultrashort pulses launched into ChG fibers with different core diameters *d* and tellurium content in the core glass. It is well known that not only fiber nonlinearity, depending on $n_2$ and effective mode field area, impacts on SC generation, but dispersion is also of great importance. The ChGs are chosen with a large refractive index difference ($\Delta n$) between them. This allows tailoring dispersion by core diameter variation due to the strong waveguide contribution. The content of Te in As-Se-Te glasses affects refractive index, and can be also varied for dispersion management. Besides, As-Se-Te glasses have a number of advantages over binary arsenic chalcogenides. They offer lower phonon energies, higher values of refractive index (2.9-3), higher nonlinear refractive indices ($n_2$), better transmission in the long wavelength mid-IR range, and high stability against crystallization. So, we believe, that the proposed fiber design is suitable for ultra-broadband spectral conversion due to the above listed advantages of physical and chemical properties.

**Results**
*Experimental*
We have synthesized high-purity As-Se-Te and As-S glasses by using physicochemical principles of the preparation developed by us previously [2,16,17].

The $As_{39.4}S_{60.6}$ glass has been prepared by direct melting of mixtures of extra pure $As_4S_4$ and S in an evacuated silica-glass ampoule. To reduce the content of impurities of oxygen, carbon and heterophase inclusions, the starting substances have been purified by chemical and distillation methods. The batches of $As_4S_4$ and sulfur have been loaded into the synthesis reactor by evaporation from intermediate ampoules. The evacuated ampoule with the charge has been placed into the rocking furnace. Then, the ampoule has been heated up to $750^{\circ}C$ for 5 h and experienced an 8 h dwell at this temperature. The melt has been solidified by air-quenching from a temperature of $400^{\circ}C$, with subsequent glass annealing at $220^{\circ}C$ for 1 h and slow cooling to room temperature. The obtained $As_{39.4}S_{60.6}$ glass has been shaped as solid rods with a diameter of 40 mm and length of 130-140 mm. The samples have been analyzed by laser mass spectroscopy to determine the impurity content of metals and silicon, and by IR-spectroscopy to determine the content of oxygen, hydrogen and carbon impurities. The impurity content in the glass is the following: hydrogen <0.1 ppm(wt), carbon <0.02 ppm(wt), silicon <0.4 ppm (wt), transition metals <0.1 ppm (wt). The absorption spectrum of the glass (Fig. 1 (a)) is characterized by the absence of intense impurity absorption bands in the 5000-1500 $cm^{-1}$ spectral range.

The $As_{39.4}Se_{55.3}Te_{5.3}$ glass has been obtained by melting the extra pure initial elements (with a purity of 6N) in an evacuated silica ampoule (from Heraeus glass) using a chemical distillation purification of glass-forming melt with chemical getters Al (700 ppm) and TeCl4 (2000 ppm) for binding and removal of hydrogen and oxygen impurities. The melt has been double distilled in an open vacuum system and then double distilled in a closed vacuum system. At the final glass synthesis stage, the melt has been homogenized at a temperature of $750\ ^{\circ}C$ for 7 hours, with subsequent air-quenching, annealing at $175^{\circ}C$ for 1 h, and slow cooling to room temperature. The absorption spectrum of the obtained $As_{39.4}Se_{55.3}Te_{5.3}$ glass is presented in Fig. 1 (b).

The content of impurities of hydrogen and oxygen measured by IR-spectroscopy is 0.02 and 0.1 ppm(wt), respectively. Table 1 lists the content of impurities of As-Se-Te glass measured by the chemical atomic emission spectroscopy.

To measure the material optical loss of the $As_{39.4}Se_{55.3}Te_{5.3}$ core glass, the $As_{39.4}Se_{55.3}Te_{5.3}/As_{38}Se_{62}$ core/clad multimode fiber has been fabricated by the double crucible method. High purity $As_{38}Se_{62}$ glass has been prepared analogously using the chemical and distillation purification. The fiber core diameter is 300 μm, and the clad diameter is 400 μm. A standard cut-back technique is implemented to measure optical loss. The minimal optical loss of 0.065±0.005 dB/m at a wavelength of 6.3 μm (Fig. 2 (a)) indicates to a high purity of core glass. There are absorption bands of hydrogen with maxima at 3.53 μm, 4.12 μm, and 4.57 μm in the spectrum. The optical loss of the most intensive Se–H band is 2.2 dB/m at 4.57 μm. This $As_{39.4}Se_{55.3}Te_{5.3}$ glass has been used to fabricate a core of single-mode fibers.

Fabrication of optical fibers with core diameters of 0.5-1.7 μm has been carried out in 3 stages:

1. $As_{39.4}Se_{55.3}Te_{5.3}/As_{39.4}S_{60.4}$ step-index fibers with core diameters of 9-11 μm and clad diameters of 260-264 μm have been drawn by the double crucible method (Fig. 3(a)).

2. $As_{39.4}S_{60.4}$ glass capillaries have been drawn with inner diameters of 280-300 μm and outer diameters of 2.7 mm.

3. The $As_{39.4}Se_{55.3}Te_{5.3}/As_{39.4}S_{60.4}$ step-index fiber produced at the first stage has been placed inside the $As_{39.4}S_{60.4}$ capillary and drawn by the "rod-in tube" method to the thinner fiber. During drawing process, the interfacial space between the core rod and the clad glass tube has been evacuated down to 300 mm Hg, to collapse the cladding tube onto the rod.

More than 20 meters of fibers with outer diameters of 240-320 μm and core diameters of 0.5-1.7 μm have been drawn (Fig. 3(b)). For fiber protection and prevention of rapid aging, coating of a fluoroplastic, F-42, is used.

*Theoretical*

We evaluate group velocity dispersion and effective mode field area for the fundamental mode $HE_{11}$ of As-Se-Te/As-S core/cladding fibers for different core diameters and tellurium concentration. We use a standard procedure for finding electric and magnetic field components and wavenumber $\beta$ as the exact solution of Maxwell's equation for step index profiles [18]. The corresponding characteristic equation [18] is solved numerically by the Newton method. Refractive index measurements of $As_2S_3$ and $As_2Se_3$ by Amorphous Materials Inc. ("AMTIR-6" and "AMTIR-2") [19] are taken as estimates for our $As_{39.4}S_{60.6}$ and $As_{39.4}Se_{60.6}$ glasses respectively. We assume that the contribution of tellurium in the refractive index $n$ of $As_{40}Se_{60-x}Te_x$ is proportional to $x$:
$n(As_{40}Se_{60-x}Te_x) = n(As_2Se_3) + 0.013 \cdot x$ [20]. The calculated dispersion $D = -2\pi c(\partial^2\beta/\partial\omega^2)/\lambda^2$ (where $c$ is the speed of light in vacuum, $\lambda$ is the wavelength in vacuum, $\omega$ is the angular frequency) is shown in Fig. 4 for different $d$ and $x$. The effective mode field area defined as $A_{eff} = (\int S_z d^2\mathbf{r})^2/(\int S_z^2 d^2\mathbf{r})$ (where $S_z$ is the longitudinal component of the Poynting vector and $\mathbf{r}$ is the cross sectional position vector) is also depicted in Fig. 4. One can see that the dispersion strongly depends on both, $d$ and $x$. The dispersion is all-normal in the spectral range of interest for relatively low tellurium concentrations ($x \leq 10\%$) and thin cores ($d \leq 2$ μm). Two zero dispersion wavelengths (ZDWs) appear with anomalous dispersion between them with increasing tellurium concentration. But for core diameters of about 5 μm and thicker, the fiber dispersion is approaching the material dispersion having only one ZDW in the mid-IR. $A_{eff}$ increases monotonically with growing wavelength. For $\lambda/n > d$, the mode field is significantly worse localized near the core than for $\lambda/n < d$, so $A_{eff}$ enlarges very fast for $\lambda > d/n$.

To simulate pulse propagation leading to SC generation we employ the generalized nonlinear Schrödinger equation with the calculated dispersion taken into account, the third-order instantaneous Kerr and retarded Raman nonlinearities, linear loss, as well as frequency dependence of effective mode field area [21,22] with the addition of two-photon absorption (TPA). Constants and functional dependences are taken from the reported data for $As_2Se_3$ glasses and fibers. So, we use: $n_2 = 2.2 \cdot 10^{-17}$ m$^2$/W [23]; the approximation of the Raman response function by a two decaying harmonic oscillators function [24]; and the TPA coefficient dependence on frequency as in [12]. The nonlinear coefficient is $\gamma = 2\pi n_2/(\lambda A_{eff})$. The split-step Fourier method using the fast Fourier transform is applied to solve an unidirectional nonlinear field propagation equation. Although the fibers are multimode at 2 μm, we simulate pulse propagation only for the fundamental mode $HE_{11}$ in accordance with the work [25] demonstrating that, if the pump pulse is short enough ($\leq 10$ps), higher order modes are not important for SC bandwidth because of temporal walk-off; no light is transferred between the modes [25].

Figure 4 demonstrates the simulated SC spectra at the output of 2-cm long ChG fibers with different core diameters and tellurium concentrations. The input pulse energy is 100 pJ, the pulse duration is 150 fs, and the central wavelength is 2μm. One can see that for all analyzed compositions of core glasses, the maximum spectral broadening is observed for $d = 2$ μm. The optimal tellurium content is about 30% and higher when the pump wavelength is located near the ZDW. In this case, the red SC boundary is beyond 8 μm. For $x \leq 10\%$, the SC generation occurs in the all-normal dispersion regime and the maximum expected wavelength is about 4 μm. For $d \geq 5$ μm and each $x$, the pulse is broadening fast in the time domain due to the large dispersion. Its peak intensity is quickly reduces along the propagation distance, and spectral broadening determined by self-phase modulation is not large.

Further, we simulate nonlinear dynamics of signals propagating in the optimal fiber with 2-μm core of $As_{40}Se_{30}Te_{30}$ glass. The spectral evolution is presented in Fig. 5 (left column) for different energies. The simulated spectra after a 1-cm fiber, demonstrating wavelength conversion beyond 8 μm for energies exceeding 50 pJ, are shown in Fig. 5 (right column).

**Discussion**

We propose and theoretically optimize a simple design of a mid-IR laser source for generating SC extending from ~1 μm to more than 8 μm. It is based on using an As-Se-Te/As-S core/cladding step-index fiber with two ZDWs and all-fiber femtosecond laser system at 2 μm. To the best of our knowledge, the possibility of spectral broadening beyond 8 μm of femtosecond optical pulses at the wavelength of 2 μm in ChG fibers is demonstrated for the first time. Ultra-broadband spectra extending more than two octaves are demonstrated numerically after only a few millimeters of the $As_{40}Se_{30}Te_{30}/As_{40}S_{60}$ step-index fiber with a very small core diameter of 2 μm for pump energy of about 50 pJ and higher. This becomes possible because the spectrum initially located in the normal dispersion region is expanding fast beyond the first ZDW and part of the energy found in the anomalous dispersion region is sufficient for soliton formation. After that, soliton self-frequency shift due to simulated Raman scattering is observed. As the Raman soliton central wavelength approaches the second ZDW, the red-shifted dispersive wave generation occurs in the normal dispersion region beyond the second ZDW [10]. An increase in pump energy over 100 pJ does not lead to significant SC red boundary expanding beyond 8 μm due to high normal dispersion and rapidly decreasing nonlinearity. The SC blue boundary of about 1 μm is limited by TPA.

The earlier reported experimental and theoretical works demonstrating SC generation with the same or longer red boundary are based on using mid-IR pump sources such as OPA or SC generated with fluoride fibers. Using for pumping an all-fiber system based on standard telecom components and Er-doped and/or Tm-doped silica fibers simplifies significantly the scheme of SC sources and may make them more convenient for applications.

It should be emphasized that, in application to the problem of the development of coherent SC laser sources, preparation of high-purity glasses with proper physical and chemical characteristics, geometrical design and manufacturing fibers with tailoring dispersion, and optimization pumping scheme are very essential and should be considered jointly. We have also synthesized high-purity $As_{39.4}Se_{55.3}Te_{5.3}$ glass which is characterized by excellent transmittance in the mid-IR range and minimum material optical loss of 0.065±0.005 dB/m at a wavelength of 6.3 μm. Then, single-mode $As_{39.4}Se_{55.3}Te_{5.3}/As_{39.4}S_{60.6}$ glass fibers with core diameter within 0.5-1.7 μm are fabricated. In our forthcoming experimental study, we are planning to implement mid-IR SC generation with created single-mode $As_{39.4}Se_{55.3}Te_{5.3}/As_{39.4}S_{60.6}$ glass fibers as well as to fabricate fibers with a different composition of core glass.

So, we believe that the proposed design of SC laser source can be useful for the development of mid-IR fiber optical systems and ChG-based photonic devices.

**Conclusion**

An optimal mid-IR SC laser source based on As-Se-Te/As-S core/clad step-index fibers and a femtosecond all-fiber laser system at 2 μm has been proposed and demonstrated theoretically. Numerically simulated spectra have been shown to extend from ~1 μm to more than 8 μm for pump energy of order 100 pJ in a fiber with a core diameter of 2 μm. To the best of our knowledge, the possibility of such long-wavelength spectral conversion of pump pulses at the wavelength of 2 μm in optical fibers has been demonstrated for the first time. The theoretical calculations have been performed on the basis of real low loss step-index As-Se-Te/As-S glass fibers with core diameter within 0.5-1.7 μm.

**Acknowledgments**

The experimental work is supported by the Russian Science Foundation (Grant No. 16-13-10251). The numerical simulation provided by E.A.A. is supported by the Russian Foundation for Basic Research (Grant No 16-32-60053).

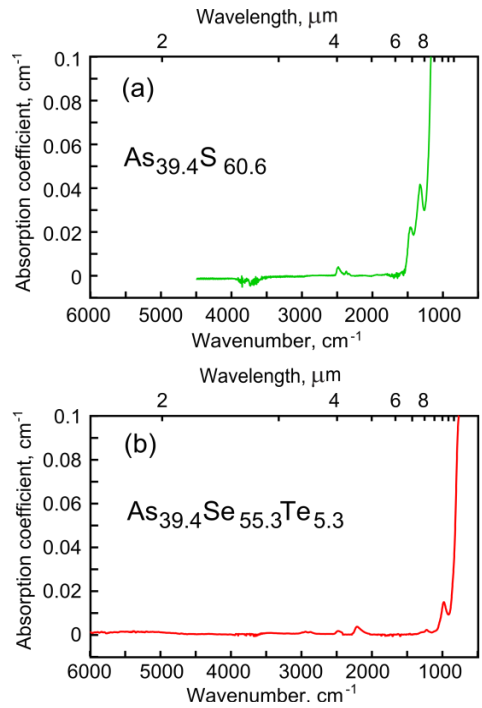

Fig. 1. Absorption spectra of glasses: (a) $As_{39.4}S_{60.6}$ (optical path is 145 mm); (b) $As_{39.4}Se_{55.3}Te_{5.3}$ (optical path is 115 mm).

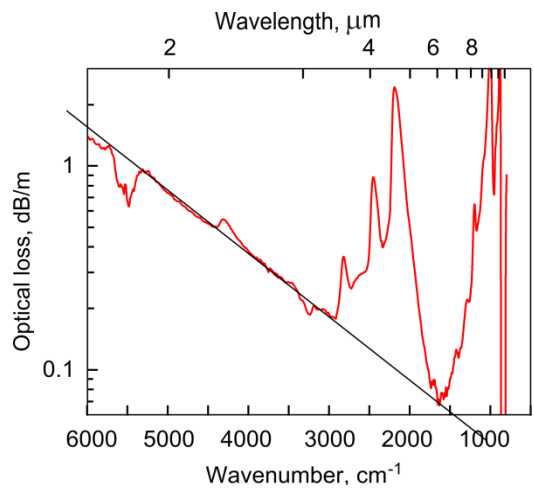

Fig. 2. Spectrum of total optical loss of $As_{39.4}Se_{55.3}Te_{5.3}/As_{38}Se_{62}$ core/clad glass fiber

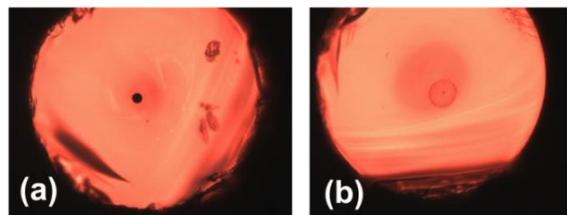

Fig. 3. Cross sections of $As_{39.4}Se_{55.3}Te_{5.3}/As_{39.4}S_{60.4}$ fibers: (a) drawn from melts by double crucible technique with core diameter of 11 μm and clad diameter of 264 μm; (b) drawn by "rod-in-tube" technique with core diameter of 1.4 μm and clad diameter of 320 μm.

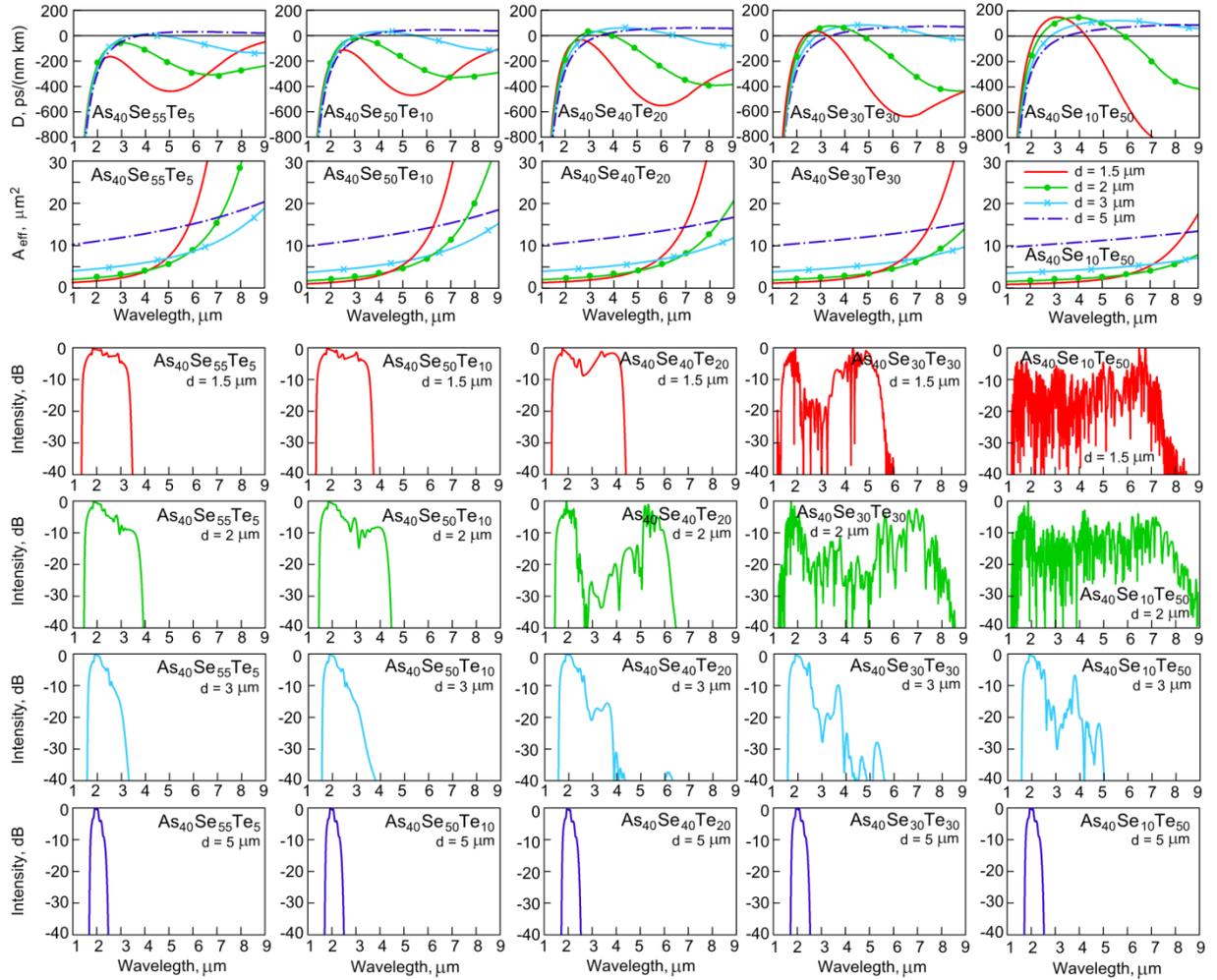

*Fig. 4. Numerically calculated: dispersion of the fibers with various tellurium content in the As-Se-Te core (the top row), effective mode field areas (the second row); spectra at the output of 2-cm long fiber pumped by 150-fs pulses at 2 μm with energy of 100 pJ (the 3rd, 4th, 5th, and 6th rows).*

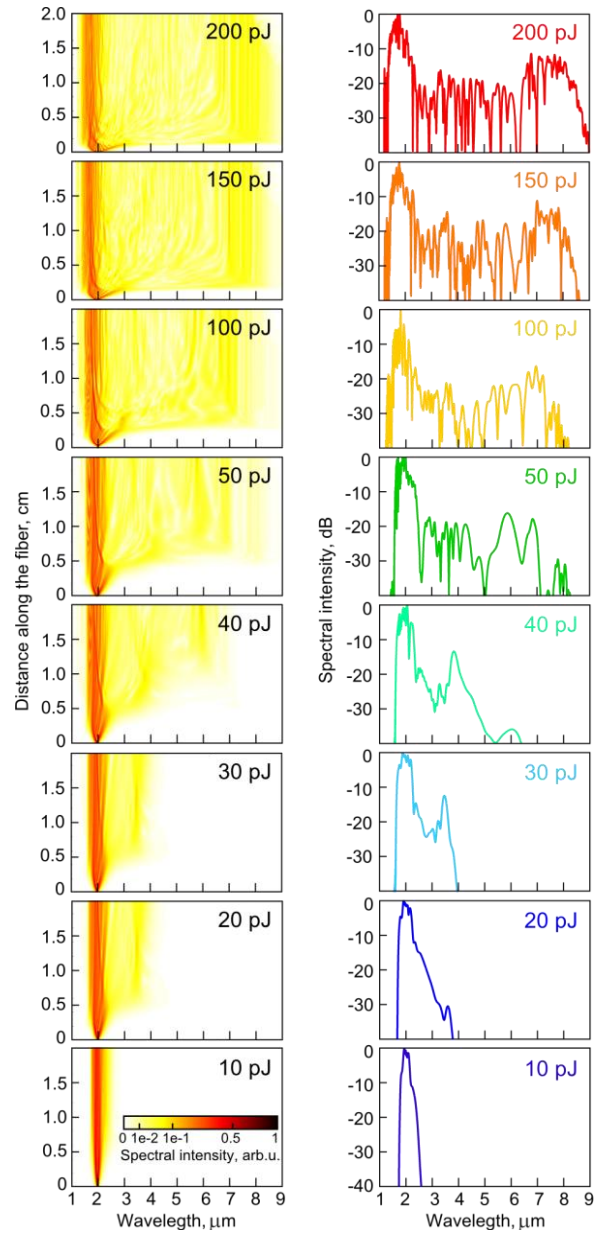

*Fig. 5. Numerically calculated spectral evolution of 150-fs pulses with different energies in $As_{40}Se_{30}Te_{30}/As_{40}S_{60}$ step-index fiber having core diameter of 2 µm (left column). Spectra after 1 cm propagation in this fiber (right column).*

*Table 1. The content of impurities in $As_{39.4}Se_{55.3}Te_{5.3}$ glass*

| Impurity | Content, ppm(wt) | Impurity | Content, ppm(wt) |
|---|---|---|---|
| B | <0.01 | Cr | <0.1 |
| C | 0.5 | Mn | <0.07 |
| N | 0.09 | Fe | <0.1 |
| F | 0.8 | Co | <0.07 |
| Na | 0.02 | Ni | <0.1 |
| Mg | 0.03 | Cu | <0.1 |
| Al | 0.3 | Zn | <0.2 |
| Si | 0.1 | Ga | <0.3 |
| P | <0.04 | Ge | <0.5 |
| S | ≤0.08 | Ag | <0.5 |
| Cl | <0.04 | Cd | <0.8 |
| Ca | <2 | In | <0.2 |
| Sc | <0.06 | Sn | <0.8 |
| Ti | <0.1 | Sb | <0.5 |
| V | <0.06 | I | <0.2 |